# A nonlinear capillarity-driven grain growth in polycrystalline materials

**Authors:** Jianfeng Hu[1]*, Xianhao Wang[2]

**Affiliations:**

[1] School of Materials Science and Engineering, Shanghai University, Shanghai 200444, China.

[2] Carl Zeiss Co., Shanghai, 60 Meiyue road, Shanghai 200131, China.

*Correspondence to: jianfenghu0611@gmail.com or jianfenghu@shu.edu.cn.

**Abstract**: A formula of grain growth rate, based on a nonlinear capillarity-driven relation, is derived to predict and interpret realistic growth processes in polycrystalline systems. The derived formula reveals how the growth and stagnation of grains dominated by the correlated parameters (temperature, interfacial energy, step free energy, grain size and size distribution in polycrystalline system etc.). Our study provide a conclusive model of the growth and stagnation of grains, and thus offers helpful guides for the microstructural design to optimize the properties of polycrystalline materials.

## 1. Introduction:

The practical performances of polycrystalline materials are strongly affected by the formed microstructure inside, which is mostly dominated by grain-growth behaviors (*1–3*). Grain growth has been extensively investigated to understand grain-growth behavior and control the microstructure for more than 60 years [4–8]. For the capillarity-driven grain growth, the linear relation between the velocity of a grain-boundary migration and capillary driving force, *i.e.* $v = M \cdot \Delta P$, is generally assumed to model the grain growth behavior. Here ν, $M$ and $\Delta P$ are the velocity, kinetic coefficient describing its mobility and capillary driving force. Based on this assumption, the well-known von Neumann-Mullins law was derived to depict the growth of individual grains or bubbles (*4, 6, 9*). The three-dimensional von Neumann-Mullins law was given by $\frac{dV}{dt} = -2\pi M\gamma(D - \frac{1}{6}\sum_{i=1}^{n} e_i)$, which has the constant GB mobility $M$ and unified GB free energy $\gamma$ for all facets of grain (*6*). This equation shows that the growth rate of individual grain $\frac{dV}{dt}$ is linearly related with the difference between grain's linear size $D$ and the total length of its edges $e_i$. It suggests that the growth of individual grains is dominated by its surface topography and grain size. Similarly, another well-known growth law for individual grains, *i.e.* $\frac{dR}{dt} = \alpha M\gamma(\frac{1}{R_c} - \frac{1}{R})$ (here, $\alpha$, $R_c$ and $R$ are dimensionless constant, a critical grain radius and grain radius.), was proposed by Hillert (*5*). This law, based on mean-field approximation, suggests that the growth of individual grains is dominated by its linear grain size and a critical grain size associated with grain size distribution. Both von Neumann-Mullins law and Hillert's law, using the unified GB mobility and GB free energy for all grains, lead to a natural tendency toward a fixed distribution of relative sizes during grain growth. This growth behavior with unimodal grain-size distribution (uni-GSD) can be called statistical self-similarity growth, *i.e.* normal (or "ideal") grain growth (NGG) (*10–12*). However, the microstructural evolution in real polycrystalline materials was frequently observed to be abnormal (discontinuous) grain growth (AGG) corresponding to a bimodal grain-size distribution (bi-GSD) (*13*). Besides AGG, the kinetics behavior of grain-growth stagnation (GGS) in polycrystalline materials cannot be interpreted by von Neumann-Mullins law or Hillert's law.

Recently, numerous studies have indicated the close correlation between GB features (including structures and chemical compositions) with grain growth behaviors in polycrystalline systems. Many mechanisms or models have been proposed to explain AGG behavior(*14–20*). The recent models believed that solute drag effect or particle/pore pinning at GBs might result in AGG (*15*). More recently, Harmer et al.(*18*) proposed that the coexistence of various GB complexions, corresponding to the different GB mobilities, resulted in the occurrence of AGG. Meanwhile, Kang at al.(*19*) proposed a mixed control of boundary migration model that grain growth behaviors determined by the relationship between the maximum driving force and a presumptively critical driving force in polycrystalline materials. According to this model, AGG occurred in the case of the maximum driving force slightly higher than a critical driving force. In contrast to the unified mobility assumed in early classical laws, these recent models mostly attribute grain growth behaviors to different GB mobilities associated with the complicated GB features, e.g. AGG attributed to the coexistence of various types of GB mobilities. Each model is valid in certain regimes. NGG and AGG are the statistical growth behaviors of individual grains in polycrystalline materials, which may result from the same growth rule but different growth rates of individual grains that strongly influenced by GB features. However, due to the highly complex nature of grain boundaries, a validated mathematic relationship between the growth rate of individual grains and GB-related parameters has yet to be established for grain growth despite significant progress in recent theory studies(*12*, *21*, *22*). Therefore, interpreting theoretically grain growth behaviors of NGG and AGG has long remained elusive due to the lack of a conclusive growth model of individual grains to depict the true picture of grain growth in polycrystalline system.

Here, we derives a growth formula of individual grain from a nonlinear relationship. The derived formula incorporates the important factors influencing grain growth behaviors, such as GB-related parameters (GB step free energy and GB energy) and GSD-related parameters. This formula reveals how the growth and stagnation of individual grains are dominated by the correlated physical parameters, especially GB-related parameters. This formula offers a conclusive model for the quantitative study on grain growth behavior and for the design of polycrystalline materials

## 2. Results and discussion

We suppose that growth rate of grain is mostly determined by the two-dimensional nucleation rate of atoms at the surface of growing grain, since the recent experimental and theoretical results indicated a liquid-like film occurring at surface or interface of grains during sintering (*23*). The driving force results from the surface curvature difference between adjacent grains on both sides of the GB. Analogous to the classical crystal growth theory (*24*), we can obtain the growth rate of $i_{th}$ face of individual polyhedron grains in polycrystalline materials as follows:

$$v_i = (Jv^2)^{1/3} h \cdot exp\left(-\frac{\Delta G_i^*}{3kT}\right) = C \cdot exp\left(-\frac{\Delta G_i^*}{3kT}\right) \qquad 1$$

Here, $J$ and $v$ are the pre-exponential factor of nucleation and the rate of the step advance, respectively. $h$, $k$ and $T$ are the step height, the Boltzmann's constant is and the absolute temperature. $\Delta G_i^*$ is the critical energy barrier to form a two-dimension nucleus on the surface of growing grains. Although theoretical estimations of pre-exponential factors of $C = (Jv^2)^{1/3} h$ are generally uncertain, their values are essentially insensitive to small changes of temperature and thus can be usually treated as approximate constants. To form a disk-like nucleus of height $h$ and radius $R$ on the $i_{th}$ face, the free energy change is given by $\Delta G_i = 2\pi R \varepsilon_i - \pi R^2 h \cdot \Delta P_i$. Therefore, the work needed to form the critical 2D nucleus is obtained as $\Delta G_i^* = \pi \varepsilon_i^2 / (h \cdot \Delta P_i)$. Here $\varepsilon_i$ is the step free energy per unit length on the $i_{th}$ face. $\Delta P_i$ is the driving force to form a nucleus on the $i_{th}$ face. Further, Eq.1 can be rewritten as

$$v_i = C \cdot exp\left(-\frac{\pi \varepsilon_i^2}{3kTh \cdot \Delta P_i}\right) \qquad 2$$

For grain growth in the polycrystalline materials, the migration of grain boundary is driven by the curvature difference between the adjacent grains on both sides of a nanometer-thick boundary. The mean curvatures of smaller grains are statistically larger than that of larger grain, since grain growth has been classically described in terms of growth of larger grains at the expense of smaller grains, *i.e.* Ostwald ripening. The driving force $\Delta P_i$ for the growth of the $i_{th}$ face is given by $\Delta P_i = 2\gamma_i(\kappa_{ai} - \kappa_i)$ (here $\gamma_i$ is the interfacial free energy associated with the $i_{th}$ face. $\kappa_i$ and $\kappa_{ai}$ are the mean curvatures of the $i_{th}$ face of growing grain and the adjacent face of the surrounding smaller grain, respectively.) Therefore, the growth rate on the $i_{th}$ face of polyhedron grain in Eq. 2 can be rewritten as follows.

$$v_i = C \cdot exp\left(-\frac{\pi \varepsilon_i^2}{6kTh\gamma_i \cdot (\kappa_{ai}-\kappa_i)}\right) = C \cdot exp\left(-\frac{\pi \varepsilon_i^*}{6kT \cdot \kappa_i(n_i-1)}\right) \qquad 3$$

where, $\varepsilon_i^*$ and $n_i$ are equal to $\varepsilon_i^2/h\gamma_i$ and $\kappa_{ai}/\kappa_i$, respectively. $\varepsilon_i^*$ can be named the effective step free energy due to the same dimension with $\varepsilon_i$. Analogous to $\varepsilon_i$ (25, 26), $\varepsilon_i^*$ is also a temperature-dependent variable but more sensitive to the temperature than $\varepsilon_i$. According to the definition, the dimensionless parameter $n_i$ is determined by the topographic surfaces of polyhedral grains between both sides of the GB that closely associated with the corresponding grain sizes. The effect of $n_i$ on growth rate is statistically limited by GSD in polycrystalline materials. Finally, the growth rate of the volume $V$ of individual grain (polyhedron with n faces) can be expressed as

$$\frac{dV}{dt} = \sum_{i=1}^{n} v_i \cdot A_i = \sum_{i=1}^{n} CA_i exp(\frac{-\pi \varepsilon_i^*}{6kT \cdot \kappa_i (n_i - 1)}) \qquad 4$$

where $A_i$ the area of $i_{th}$ face of growing polyhedron grain. In contrast to the von Neumann-Mullins relation and Hillert's law, there is a nonlinear relation between capillary driving force and the GB migration rate. The formula in Eq. 4 reveals that the growth of individual grains exponentially depends on the reciprocal of grain curvature and the effective step free energy for grain growth. According to the definition, the effective step free energy $\varepsilon^*$ is jointly determined by the crystallographic features of growing grain and its surrounding GB conditions (e.g. chemical compositions and GB structures) in addition to temperature. In a word, the derived formula reflects the effects of GB features and GSD on the growth rate of individual grains, which is consistent with the experimental observations and characterizations.

**Conclusions**

The derived formula well reveals how the growth and stagnation of individual grain are dominated by the combination of grain size and the effective step free energy at GB. It also demonstrates that AGG occurs even though a single $\varepsilon^*$ value (one type of boundary features) dominating the growth of individual grains, which is distinctly different from the above-mentioned interpretation for AGG in recent models. Furthermore, according to Eq. 4, the anisotropic $\varepsilon^*$ values existing at the surface of growing grains may lead to anisotropic grain growth and then form the anisotropic morphology of grains such as the rod-like $Si_3N_4$ and plate-like $Al_2O_3$ in polycrystalline materials. On the other hand, the coexistence of various $\varepsilon^*$ values may result in a statistically continuous grain growth, *i.e.* NGG, in polycrystalline materials. In a word, the Eq. 4 offers a conclusive model for exactly interpreting and predicting the grain growth behaviors and microstructural evolution in polycrystalline materials. Furthermore, it may allow us to accurately tailor and design properties of polycrystalline materials.